\input{aipcheck}

\documentclass[
    ,final            
  ]
  {aipproc}

\layoutstyle{6x9}

\begin{document}

\title{Could a foliation by constant mean curvature hypersurfaces cover the existence of most observers in our part of spacetime?
}


\author{Prado Martin-Moruno}{
  address={School of Mathematics, Statistics, and Operations Research,\\
Victoria University of Wellington, PO Box 600, Wellington 6140, New Zealand}
}

\begin{abstract}
We present a foliation by constant mean curvature hypersurfaces of a de Sitter space with a thin-wall Coleman-De Luccia bubble of de Sitter space inside, which covers the existence of most observers in our part of spacetime if we are placed in the region outside the bubble.

\end{abstract}

\maketitle


If the current accelerated expansion of our universe is due to a cosmological constant, then our future Universe will be a de Sitter space where a bubble containing another universe could nucleate.
That nucleation process, first studied by Coleman and De Luccia (CDL) \cite{Coleman}, takes place when considering the cosmological constant as a vacuum energy, corresponding to a minimum value of a particular potential.  If this potential has two minima, then the false vacuum could decay into the true vacuum.
An observer placed in the original de Sitter space would interpret such a process as the nucleation of a bubble growing at a velocity close to the speed of light, destroying part of his/her universe.

On the other hand, a de Sitter space with a CDL bubble of de Sitter space inside is compound of two infinite spaces; thus, anything that can occur with non-vanishing probability will take place.
To assign some probability to the occurrence of different phenomena one should, first of all, count the events of that spacetime.
Page has suggested \cite{Page} that constant mean curvature (CMC) hypersurfaces could have a crucial interest regarding an approach to the measure problem in eternal inflation. He has noted that if they could cover our existence, then they would provide us with at least approximately the right measure of our observations. Following this spirit, we consider a scenario with only one bubble, which nucleates due to the existence of a cosmological constant driving the acceleration of our Universe, and study whether a CMC foliation covers our existence \cite{Yo}.

The trajectory of the bubble wall in the outside space can be obtained easily using the coordinates in the 5-dimensional space  (where a de Sitter space can be visualized as a hyperboloid \cite{HyE}). Noting that due to the symmetry the trajectory of the wall in the outside space can be fixed at a value of $w=D$, and considering, without lost of generality, that the bubble nucleates at $t=0$ and centered at $\eta=0$, one can express this trajectory using the coordinates of the closed slicing as \cite{Yo}
\begin{equation}\label{trayectoria}
\eta^*(t)={\rm arccos}\left[\frac{D}{\alpha
\cosh(t/\alpha)}\right],
\end{equation}
with $D<\alpha$ for $D>0$. On the other hand, the junction conditions \cite{Israel} must be fulfilled on the bubble wall. The first one requires that the metric of the outside space evaluated on the wall must be equal to the inside metric on the wall.
This condition implies \cite{Yo} that the trajectory of the wall in the inside space takes a similar form to that of the outside space, with $D_b^2=D^2+\alpha_b^2-\alpha^2$, and that $\sinh(t_b/\alpha_b)=\alpha/\alpha_b \sinh(t/\alpha)$ on the wall.
The second junction condition relates the value of the difference of the extrinsic curvatures of both regions on the wall to the surface density of the energy momentum tensor on the wall; thus, the particular value of $D$ depends on the value of both cosmological constants and on the particular form of the potential presenting the two minima.


The hypersurfaces with CMC in the considered space have also CMC in a de Sitter space.
Therefore, in the first place, we must find CMC foliations of a de Sitter space.
Noting that a de Sitter hyperboloid, $-v^2+w^2+(x^i)^2=\alpha^2$, is a CMC hypersurface in the 5-dimensional Minkowski space, one could think \cite{Yo} that the intersection of two 4-hypersurfaces with CMC in a 5-dimensional Minkowski space produces a 3-hypersurface which would also have CMC, when considering this hypersurface defined in the spacetime given by one of the original 4-hypersurfaces.
Taking this argument and the symmetry of the problem into account one can consider the simplest 4-dimensional hypersurfaces with CMC in a Minkowski space. Those are hyperplanes which can be seen as lines in a $v-w$ section of the space, i. e. $\Sigma_4:\,\,v=b\,w+a,$ where $b$ is the slope and $a$ the $v$-intercept.
The intersection of these hyperplanes and the hyperboloid leads to
\begin{equation}\label{cosh}
\cosh(t/\alpha)=\frac{b\,a\cos\eta+\sqrt{a^2+1-b^2\cos^2\eta}}{1-b^2\cos^2\eta},
\end{equation}
which really are CMC in a de Sitter space, since $K=-3\frac{a}{\alpha\sqrt{a^2+1-b^2}}$.

In the second place, we consider that the hypersurfaces are given  by a similar expression in the inside space.
Therefore, in order to have CMC throughout the whole surface, we must require $K=K_b$, (i).
On the other hand, the surfaces must be regular on the wall. 
Therefore, they must fulfill the 1st junction condition on the wall, (ii), and have a regular orthonormal vector.
Since we cannot compare vectors in different spaces, we require that the scalar product of the orthonormal vector to the hypersurface and the orthonormal vector to the  wall at their intersection must be constant, (iii).
Thus, we have a system of three equations, (i), (ii) and (iii), with four unknown parameters. 
It can be seen \cite{Yo} that there are two sets of solutions describing two different foliations.
One of those foliations can be studied without fixing the parameters of the particular model, this is given by
\begin{eqnarray}\label{ch1}
\cosh\left(t/\alpha\right)=\frac{-b^2\cos\eta+\sqrt{b^2+\lambda^2(1-b^2\cos^2\eta)}}{\lambda(1-b^2\cos^2\eta)},
\end{eqnarray}
where $\lambda\equiv D/\alpha$, $0<\lambda<1$, in the outside region, and
\begin{eqnarray}\label{ch1b}
\cosh\left(t_b/\alpha_b\right)=
\frac{-b^2\lambda_b\cos\eta_b+\beta\sqrt{\beta^2+b^2(1-\lambda_b^2\cos^2\eta_b)}}{\beta^2-\lambda_b^2
b^2\cos^2\eta_b},
\end{eqnarray}
with $\lambda_b\equiv D_b/\alpha_b$, $0<\lambda_b<1$, and
$\beta\equiv D/\alpha_b$, in the region inside the bubble. These hypersurfaces are well-defined for values of $b$ in the interval $[-1,0]$.

\begin{figure}
  \includegraphics[height=.3\textheight]{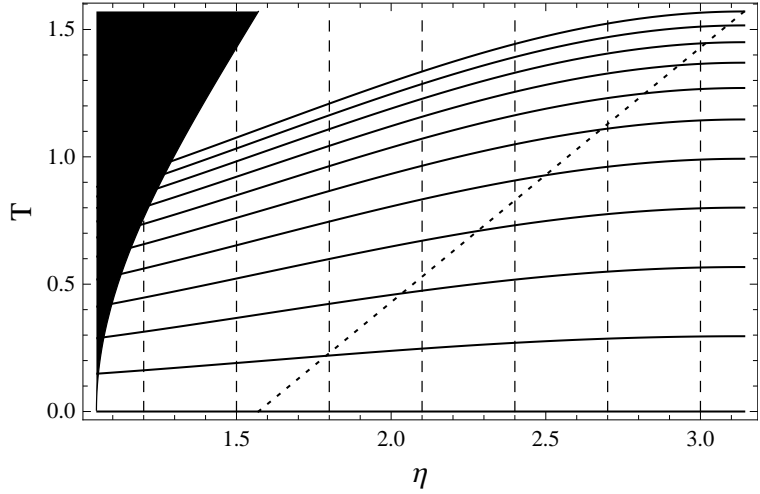}
  \caption{Region $T\geq0$ of the Carter-Penrose diagram of the outside space for a model with
$\lambda=0.5$, where the black region is not part of this space. The
closed slicing covers the whole diagram, whereas the flat slicing can
only describe the region of this diagram on the left of the dotted
line. The CMC hypersurfaces and the geodesics of the congruence correspond with the continuous curves and the dashed vertical lines, respectively.}\label{uno}
\end{figure}

Let us restrict our attention to the space outside the bubble, where we would be placed.
The foliation covers an infinite region of this space, but it also avoids an infinite part, as it is shown in Fig.~\ref{uno}.
We can understand whether the foliation covers the existence of most observers in our spacetime studying how far a particular congruence of geodesics goes in cosmic time before reaching the end of the foliation.
Therefore, we consider the congruence of geodesics orthogonal to $S_0$, given by $t=0$, which corresponds to $T=0$ (being $T$ the time of the Carter-Penrose diagram \cite{HyE}).
This congruence, $\gamma_s^\mu(t)$, has affine parameter $t$ and each value of $s$ fixes a geodesic of the congruence.
If the observers would measure $\Omega$ slightly larger than $1$, then the affine parameter of the geodesics is their cosmic time, and the spacelike hypersurfaces of constant cosmic time would have closed geometry.
It can be seen \cite{Yo} that the geodesic with $s=\pi/2$, advances until a value of the outside cosmic time being covered by the foliation which can be enormously large (tends to $1.5265\Lambda^{-1/2}$) for $\lambda\rightarrow0$ ($\lambda\rightarrow1$), and the geodesics at the boundary of the space, $s=\pi$, reaches an infinite cosmic time independently of the value of $\lambda$.
On the other hand, if $\Omega$ is just $1$, as suggested by the observational data \cite{de Bernardis}, then the observers cosmic time is that of the flat slicing \cite{HyE}, which is not geodesically complete (see Fig.~\ref{uno}).
It can be seen \cite{Yo} that the geodesic at $s=\pi$ comes into the flat slicing at the spacelike infinity, being also covered by the foliation at this point, and the geodesic at $s=\pi/2$ advances from a point in the past timelike infinity until a point corresponding to a time which can be enormously large (small but non-vanishing) for $\lambda\rightarrow0$ ($\lambda\rightarrow1$).

We can also study whether the foliation cover most part of the space by considering the Carter-Penrose diagram, Fig.~\ref{uno}.
We can roughly estimate the area being covered by the foliation in this diagram by approximating the area covered by the curves corresponding to the CMC hypersurface with $b=-1$ and the wall by the area covered by the lines which have the same endpoints.
It can be seen \cite{Yo} that the portion covered by the foliation is a decreasing function of $\lambda$, being its minimum value larger than $50\%$; thus the foliation covers most of the outside space even for large values of $\lambda$. On the other hand, we could also consider the portion of the inside space covered by the foliation, which would be a decreasing function of $\lambda_b$ and $\beta$. That portion covers most part of the inside space only for very particular decay models \cite{Yo}.

In summary, if our future Universe is a de Sitter space with a CDL bubble of de Sitter space inside, then this foliation can cover our existence. Thus, it will consistently lead, at the end of the day, to a non-vanishing probability for the occurrence of the phenomena that we measure; which would allow us to be typical observers by choosing some suitable measure.
On the other hand, the foliation also covers most part of our region of space. 
Therefore, in the cases that the total portion, considering both regions, is larger than $50\%$, the foliation can be used to approximate the events of this spacetime; and, in the other cases, the approximation would be accurate only if our region contributes more to the path integral.


\begin{theacknowledgments}
The author acknowledges financial support from a FECYT postdoctoral mobility contract of the Spanish 
Ministry of Education through National Programme No.~2008--2011.
\end{theacknowledgments}

\end{document}